\documentclass{article}
\begin{document}

\begin{center}
\centerline{\large \bf The arrow of time in quantum mechanics and in nonlinear 
optics }
\end{center}

\vspace{3 pt}
\centerline{\sl V.A.Kuz'menko\footnote{Electronic 
address: kuzmenko@triniti.ru}}

\vspace{5 pt}
\centerline{\small \it Troitsk Institute for Innovation and Fusion 
Research,}
\centerline{\small \it Troitsk, Moscow region, 142190, Russian 
Federation.}
\vspace{5 pt}
\begin{abstract}

The direct and indirect experimental proofs of a strong time invariance 
violation in optics are discussed. Time noninvariance for present day becomes 
the only real physical base for explanation the origin of the most phenomena 
in nonlinear optics. This is a good cause to introduce the time asymmetry 
into the dynamical equations of the basic laws of physics.

\vspace{5 pt}
{PACS number: 33.80.Rv, 42.50.Hz}
\end{abstract}

\vspace{12 pt}

\centerline{\bf Introduction}

\vspace{3 pt}

The idea of the arrow of time is quite natural and recognized in biology, 
chemistry and other fields [1]. However, its recognition in the physics 
unexpectedly turns out to be rather problematical [2]. The main obstacle 
here is the widespread opinion, that the basic laws of physics are time 
reversal invariant [3]. 

The experimental search of C,P,T-invariance violations is the important 
part of modern physics [4,5]. CP violation was found in the kaon decays [6]. 
In atomic physics, the scientists look for the parity nonconcervation and 
atomic electric dipole moments (as a manifestation of the time reversal 
violation) [7]. There is the field of the weak interactions.

It is surprising, but in optics (in the field of electromagnetic interactions) 
the number of direct and indirect experimental proofs of strong time 
invariance violation exists for many years. Even more surprising fact 
is that the scientific community does not ready to accept these results. 
All this results were received in the works, which were not aimed to find 
any invariance violations. Only in one case, the authors understand 
with what thing they have to do.

\vspace{7 pt}

\centerline{\bf Direct proofs}

\vspace{3 pt}

There is the work [8], where the interaction of polarized laser radiation with 
the specific non-magnetic metallic planar chiral nanostructures was studied. 
The authors believe that these experimental results unambiguously show the 
evidence of broken time reversal symmetry in such unusual object.

In other experiments, the splitting and mixing of photons were studied [9-11]. 
The process of photon mixing in this case may be reversed (when the 
frequency of formed photons is equal to the frequency of initial photons 
before the splitting) or forward (when the frequency of formed photons 
differs from the frequency of initial photons). The results of these 
experiments clearly show that the forward and reversed processes are not 
equal: the reversed process is much more efficient.

The direct way for testing the time invariance violation in optics seems to be 
very simple: we need to compare the cross-section and spectral width of the 
forward process of absorption of photons by molecules with the corresponding 
parameters of the reversed process of stimulated emission of photons. However, 
this way is rather problematic for usual objects. The main obstacle is the 
connection of the spectral widths of optical transitions and laser radiation 
with the lifetime of excited states of molecules and with the laser pulse 
width. It makes impossible to separate the measurements of forward and 
reversed processes.

However, one object with unusual combination of properties exists in optics. 
There is the wide component of lines in absorption spectrum of polyatomic 
molecules [12]. Extremely wide homogeneous spectral width in this case is 
combined with the long lifetime of excited states toward the process of 
spontaneous emission. The experiments with this object [13] inevitably 
show that the value of difference in spectral widths of the forward 
process of absorption of photon and the reversed process of stimulated 
emission exceeds five orders of magnitude. Accordingly, the cross-section 
of the reversed process exceeds the corresponding value of the forward 
process in several orders of magnitude. It is worth to point out that the 
integral cross-sections of this opposite processes must be equal because 
of equality of the Einstein's coefficients. 

The discussed above examples are the direct experimental proofs of the time 
noninvariance or inequality of forward and reversed processes in optics. 
However, these examples relate to rather exotic sides of optics. Are there any 
other manifestations of the time noninvariance exist in optics? Yes, there 
are. Moreover, the situation is so, that practically all nonlinear optics is a 
manifestation of the time noninvariance. However, such manifestations usually 
are indirect, because of the direct comparison of the parameters of forward 
and reversed processes usually are complicated. 

\vspace{7 pt}
	
\centerline{\bf Indirect evidences}

\vspace{3 pt}

Here we shall discuss as indirect evidences only two nonlinear phenomena, 
which however allow understanding and explaining the origin of many other 
effects. The first phenomenon is the adiabatic population transfer in a 
two-level quantum system during the sweeping of resonance conditions [14]. 
The essence of this phenomenon is that under action of the resonant 
radiation in two-level quantum system the so-called Rabi oscillations 
of level population are observed. The frequency of the Rabi oscillations 
increases with growth of intensity of radiation. However, if the frequency 
of radiation is changed, instead of Rabi oscillations we observe the full 
population transfer from the initial level to the opposite one. This result 
does not depend from the intensity of radiation. Therefore, the obvious 
asymmetry appears in spite of its absence in the initial conditions (the 
quantum levels and radiation transitions are assumed to be equivalent). 

Nearly sixty years ago the Bloch equations were proposed for description 
of such effects [15]. For present day the Maxwell-Bloch equations are the 
base for description the dynamics of optical transitions. This description 
is really very good. However, the problem is that the Bloch equations do 
not have any clear physical sense and the physical explanation of the 
phenomenon is absent until now. 

The concepts of longitudinal $ T_{1} $ and transverse $ T_{2} $ 
relaxation times are used in the Bloch equations. The physical sense of 
$ T_{1} $  is quite clear: it characterizes the energy relaxation process. 
In contrast, the physical origin of the transverse relaxation is unknown. 
The subsequent experiments allow specifying, that the transverse relaxation 
may be of two kinds: $ T_{2}^{'} $ and $ T_{2}^{*} $ [16]. 
The $ T_{2}^{'} $ value is defined as a time of a free polarization 
decay or incoherent decay of every dipole moment of molecules. This 
process should be irreversible. The $ T_{2}^{*} $ value is defined as a 
dephasing time or as a time of a free induction decay. This process 
may be reversible in the photon (spin) echo phenomenon. However, any 
clear physical explanation of origin of   $ T_{2}^{'} $ and  $ T_{2}^{*} $
processes is absent until now. 

Such situation seems to be rather typical for the physics. Mathematicians 
create equations and models for description the physical phenomena much 
more quickly than physicists are able to understand and explain its sense. 
For example, the physicists complain about that, "the standard model 
fails to provide a deep explanation for the physics that it describes" [7]. 
The creation by D.I.Mendeleev in 1869 the Periodic law (table) may be 
also used as a good illustration here of such situation. He builds the 
periodic table of elements (the main law in chemistry) on the base of 
known its atomic weights. It allows to predict the discovery of new 
elements and to describe its properties. However, any physical sense in 
connection between the atomic weight and the chemical properties is absent. 
The physical explanation came several decades later, when the quantum 
mechanics was created. Now we know, that the chemical properties of the 
element are depended on the structure of electronic levels and on the 
nuclear charge, which only correlates with the atomic weight.
Similarly, we only now can give the physical explanation of the concept of 
transverse relaxation and origin of number of phenomena in nonlinear optics. 
It demands to accept the idea of time noninvariance or inequality of 
forward and reversed processes in optics. In order to understand what is 
the most usual manifestation of this inequality we shall examine other 
indirect evidence. 

There are the experiments of a four photon mixing in the so-called BOXCAR 
arrangement [17, 18]. The simplified scheme of such experiments is shown in 
Fig.1 for the most common case. Three different directed laser beams intersect 
at one point. Its radiation transfer the molecules from the initial level 
{\bf a} on the quantum level {\bf d}. Then the directed spontaneous 
superfluorescence appears and the molecules return into the initial state. 
The direction of the superfluorescence does not coincide with the directions 
of the laser beams and it is easily separated. The dependences of the 
superfluorescence intensity from the delays between the laser pulses allow 
studying the dynamics of vibrational and rotational motions of molecules on 
the different quantum levels {\bf b,c}. The superfluorescence is the 
consequence of the final stage of the transition of the quantum system 
into the initial state. This transition characterizes the properties of the 
reversed processes: extremely high cross-section of the optical transition 
and its space anisotropy. Therefore, such experiments are the indirect proof 
of radical difference of the reversed optical transitions from the forward 
one: the existence of space and phase anisotropy of the reversed transition. 
For the forward transition such anisotropy is not characteristic.

This conclusion may be drawn from the analysis of absorption spectrum of 
molecules. The lines in rotational absorption spectrum are very narrow. The 
value of absorption line frequency allows to calculate the molecule moment of 
inertia with the precision $\sim 0,0001 \% $. From other side the change 
of moment of inertia due to vibration of atoms even at the zero vibrational 
level usually exceeds $ 1 \% $ [19]. Consequently, the phase of vibrational 
motion does not manifest itself in the absorption spectrum and we deal with 
some averaged moment of inertia. This situation looks like to the other one, 
which theorists call as the inability to measure the absolute phase of an 
electromagnetic field [20, 21]. Apparently, the same situation exists also 
with the orientation of molecules: there is absent manifestation in 
absorption spectrum of a freely rotated molecules of any dependence of 
absorption cross-section from the orientation of molecules toward the 
direction of radiation beam. In spite of the rather common opinion that 
such dependence should exist [22]. Therefore, some processes of averaging 
in the vibrational and rotational motions of molecules exist. 

However, for the reversed transitions such process of averaging, obviously, is 
absent. In this case, the cross-section of transition into the initial state 
with the same phase of vibration and orientation in space is extremely 
large and strongly differs from the cross-section of transition into the 
state with another orientation. This situation is shown in a diagram form 
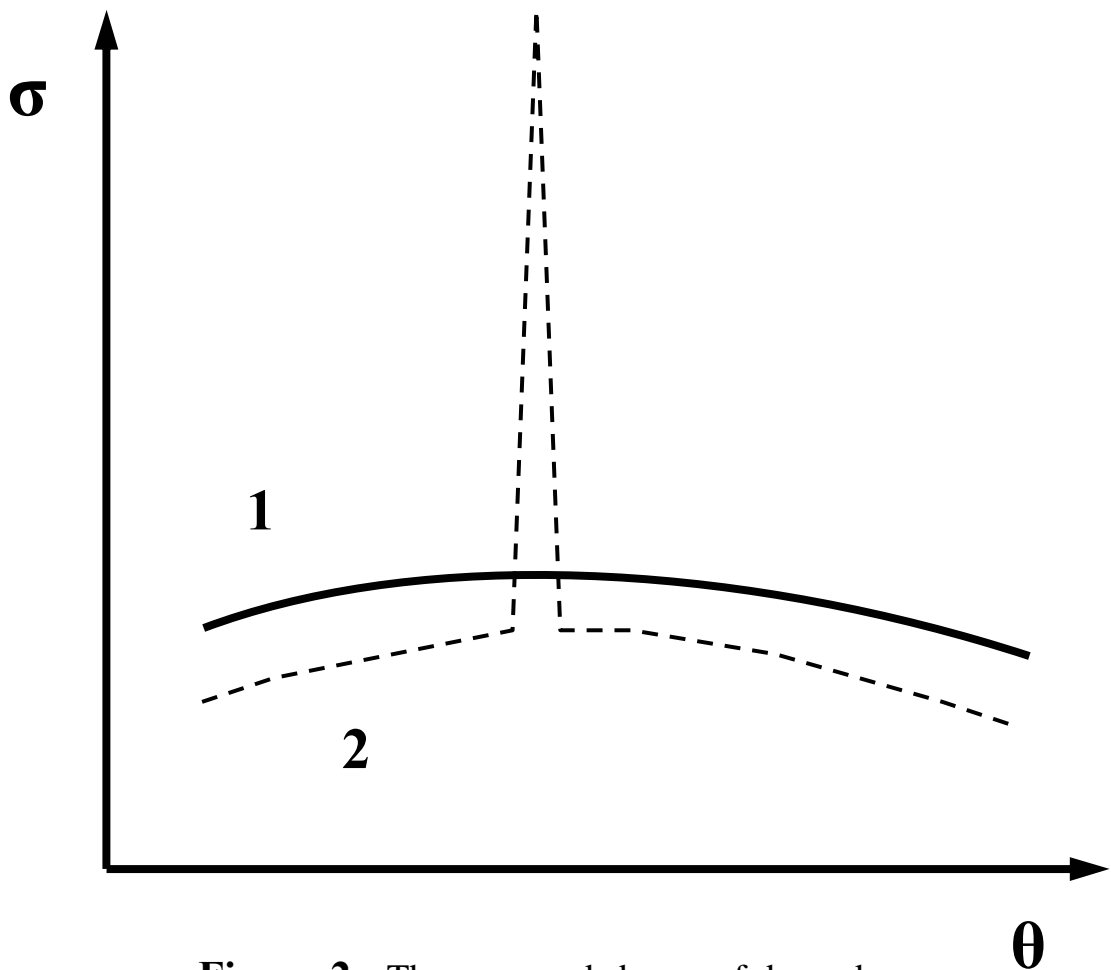
in Fig.2 as the supposed shapes of dependences of cross-section from the 
angel between the molecule axis and the direction of the laser beam. Such 
dependences, probably, may be obtained on the existent apparatus in the 
experiments with oriented in the external electric field molecules [23]. 
However, we believe, that in the present case the indirect evidences, which 
follow from the four-photon mixing experiments are also quite convincing 
proofs. 

\vspace{7 pt}

\centerline{\bf Physical explanation }

\vspace{3 pt}

In such a way the orientational and phase (vibrational) anisotropy of the 
reversed optical transitions is, obviously, most widespread and important 
manifestation of the time noninvariance in the nonlinear optics. On this base, 
we can give concrete interpretation of the origin of the transverse relaxation 
process and the origin of number of nonlinear phenomena. Thus, the 
$ T_{2}^{'} $ value characterizes the loss of memory of two-level quantum 
system about the initial state due to molecular collisions. In this case, 
the curve 2 in Fig.2 turns into the curve 1 and any difference between the 
forward and reversed transitions vanishes. Such process is irreversible.

The $ T_{2}^{*} $  value characterizes the loss of resonance for reversed 
transition due to rotation of molecules. It is defined by the average 
speed of rotation and the width of the spike on the curve 2. This process 
is periodical and reversible. After the whole period of rotation, the 
molecule turns out to be again in the resonance conditions for reversed 
transition. There is the explanation of the nutation effect.

It is easy now to explain the origin of the adiabatic population transfer 
effect. The rotation (precession) leads away the molecules from the resonance 
for the reversed transition. In this case, the cross-section of the backward 
transition on the initial quantum level turns out to be smaller, than the 
cross-section of forward transition. As a result, the molecules "get stuck" 
at the opposite quantum level. 

Furthermore, the photon (spin) echo effect can be explained as a delayed four 
photon mixing process [24]. The first laser pulse leads to absorption of one 
photon. The photon's spin transfers to molecule the rotational moment. Due to 
inhomogeneity of rotational motion, some dephasing process takes place. In the 
nuclear magnetic resonance experiments, the special inhomogeneous magnetic 
field is used for better observation of the spin echo effect. The delayed 
at time {\bf t} second laser pulse first interrupts the rotation of 
molecules by the way of stimulated emission of the photon with the same 
spin. Then the absorption of the photon with the opposite spin leads to 
the rotation of molecule in the opposite direction. After time {\bf 2t} 
the same inhomogeneity of rotational motion leads to the rephasing process and 
because the cross-section of reversed transition into the initial state 
is extremely high the directed superfluorescence emission of the photon 
echo occurs. For better understanding the mechanism of a photon echo 
effect in the molecular gas, we need to study the subtle particular feature of 
inhomogeneity of rotational motion.

In a similar manner, the concept of time noninvariance may be used for 
explanation the origin of many other effects in nonlinear optics. In contrast, 
the existent explanation of origin of nonlinear phenomena does not go 
beyond of making casual mention about the interference of a coherent 
states [25]. 

There rather old debate between theorists and experimentalists exists 
about the role of the coherency concept in physics [20]. Experimentalists 
frequently have some sense of the physical object and they persist in 
opinion that the concept of coherency or synchronism is an important 
concept of physics. However, theorists cannot find for this concept any 
reliable physical base [21]. Both sides, obviously, are right. But the 
experimentalists make a mistake when they accept the concept of coherency 
as a fundamental one. The \emph {coherency} is only a consequence of a more 
fundamental concept of the \emph {initial state}. In the same way, theorists 
do not take into consideration the time noninvariance. In this case, we do 
not deal with the measuring of an absolute phase of a wave, but with the 
measuring of a comparative phase of the reversed process, which follow the 
forward one. 

So, it is clear for present day that we need to create the new mathematical 
model (alternative analog of the Bloch equations) for description the dynamics 
of optical transitions. Such model should include the time asymmetry and 
should have a clear physical sense.

\vspace{7 pt}

\centerline{\bf Conclusion}

\vspace{3 pt}

The existent experimental results inevitably show that the time noninvariance 
is the main physical foundation of nonlinear optics. The recognition of the 
time invariance violation in optics means also a full recognition of the 
concept of the arrow of time in physics. It allows and demands to introduce 
the time asymmetry in the main dynamical laws in physics.

\end{document}